\documentclass[preprint,showpacs,preprintnumbers,amsmath,amssymb,draft]{revtex4}

\usepackage{graphicx}
\usepackage{dcolumn}
\usepackage{bm}

\begin{document}


%
\title{Metal-insulator Crossover Behavior at the Surface of NiS$_2$}


\author{D.D. Sarma*, S.R. Krishnakumar}
\affiliation{Solid State and Structural Chemistry Unit, Indian
Institute of Science, Bangalore 560 012, India}
\email{sarma@sscu.iisc.ernet.in} 
\altaffiliation[also at]{Jawaharlal  Nehru  Center  for  Advanced
Scientific Research, Bangalore, India.\\}
\author{E. Weschke, C. Sch{\"u}{\ss}ler-Langeheine, Chandan
Mazumdar, L. Kilian, G. Kaindl}
\affiliation{Institut f{\"u}r Experimentalphysik, Freie Universit{\"a}t Berlin,
 D-14195 Berlin-Dahlem, Germany}
\author{K. Mamiya,$^1$  S.-I. Fujimori,$^1$ A. Fujimori,$^{1,2}$}
\affiliation{$^1$Synchrotron Radiation Research Center, Japan
Atomic Energy Research Institute, SPring-8, Mikazuki, Hyogo 679-5148,
Japan}
\affiliation{$^2$Department of Complexity Science and
Engineering, University of Tokyo, Bunkyo-ku, Tokyo 113-0033, Japan}
\author{T. Miyadai}
\affiliation{Faculty of Fine Arts, Dohto University,
Kita-Hiroshima 061-11, Japan}


\begin{abstract}

We have performed a detailed high-resolution electron spectroscopic
investigation of NiS$_2$ and related Se-substituted compounds
NiS$_{2-x}$Se$_x$, which are known to be gapped insulators in the
bulk at all temperatures. A large spectral weight at the Fermi energy
of the room temperature spectrum, in conjunction with the extreme
surface sensitivity of the experimental probe, however, suggests that
the surface layer is  metallic at 300 K. Interestingly, the evolution
of the spectral function with decreasing temperature is characterized
by a continuous depletion of the single-particle spectral weight at
the Fermi energy and the development of a gap-like structure below a
characteristic temperature, providing evidence for a metal-insulator
crossover behavior at the surfaces of NiS$_2$ and of related
compounds. These results provide a consistent description of the
unusual transport properties observed in these systems.

\end{abstract}

\vspace{0.2in}


\maketitle

\section{Introduction}
The properties of two-dimensional (2D) electron systems, particularly
in the presence of strong interaction and disorder, are a matter of
considerable interest. While the ground state of a non-interacting
disordered 2D electron system is predicted to be an
insulator~\cite{TVR}, a recent experiment involving semiconductor
inversion-layer devices has suggested a metal-insulator transition
(MIT) in such a system~\cite{PRB}. This initiated a number of
studies~\cite{various} that aim at understanding whether such a
quantum phase transition can occur in two dimensions. The question
arises if electron interaction can lead to a metallic ground state
even in a disordered 2D system. This has actually been suggested
assuming that electron interaction opposes the effects of
disorder~\cite{prl}. A recent theoretical study, however, predicts
the absence of a true MIT in a 2D system in the simultaneous presence
of disorder {\em and} interaction and proposes instead a crossover
behavior~\cite{SDS}. The very nature of the mentioned systems based
on Si MOSFET's and various semiconducting heterostructures with very
low electron densities~\cite{PRB}, however, does not allow to study
their electronic structure by photoemission (PE), a method that has
been applied successfully to MIT's in 3D systems~\cite{RMP}.

The surface of  NiS$_{2-x}$Se$_x$ ($x\leq$~0.4) represents a
well-suited system for addressing several central questions
concerning the ground-state electronic structure of 2D systems as
well as for studying the evolution of the electronic structure by PE
across a metal-insulator-like transition in a strongly correlated 2D
system. While for bulk NiS$_{2-x}$Se$_x$, with $x>$~0.4, an $x$- and
$T$-driven MIT is well known and has been the subject of several
studies~\cite{shen1,shen2,fujimit,mot1,husmann}, we focus here on
$x\le$~0.4, where remarkable transport properties have been observed:
While optical studies show that the bulk material is a gapped
insulator at all $T$~\cite{kautz}, this is not reflected in dc
resistivity data~\cite{thio1,honigres}. Near room temperature, the
resistivity exhibits an activated behavior, with a gap of
$\approx0.2$~eV, in agreement with optical conductivity/reflectivity
measurements~\cite{kautz}. Below 120~K, however, no activated
behavior is observed, and the resistivity even decreases with
decreasing $T$. On the basis of detailed transport and Hall-effect
measurements on samples with different surface-to-volume ratios, it
was concluded that this resistivity behavior is due to a metallic
surface layer with a thickness of a few unit cells~\cite{thio1}. This
is remarkable in view of the inevitable presence of disorder in real
systems that should ensure a 2D insulating ground state~\cite{TVR}.
And in fact, this is suggested by the recent observation of a steep
increase in resistivity at low temperatures~\cite{honigres},
rendering the formation of a metallic surface layer at higher
temperatures even more intriguing~\cite{thio1}.

Many questions arise from these experimental observations, e.g., how
the low-$T$ insulating phase connects to the high-$T$ metallic phase,
whether there is a real phase transition or rather an unusual
crossover behavior, and whether the surface is metallic at higher
temperatures or the transport properties have to be interpreted in a
different way. It is known that the bulk (3D) insulating behavior of
NiS$_2$ at all temperatures is due to the formation of a Mott-Hubbard
gap in the single-particle excitation spectrum driven by
electron-electron interaction, while so far there is no information
on the nature of the low-$T$ insulating phase of the 2D surface of
this system. One may ask if its insulating state is induced by
electron-electron interaction or by Anderson localization in view of
the increased effect of disorder in the lower dimension.

Here we present the results of a high-resolution PE study of the
surface electronic structure of NiS$_2$, giving evidence for a
metallic state at high temperatures as suggested by the resistivity
data. In addition, we find that the unusual transport properties of
this compound and of the related NiS$_{2-x}$Se$_x$ system, with $x$
$\leq$ 0.4, are reflected in $T$-dependent changes of the surface
electronic structure. With electron interaction and disorder driving
the system towards localization instead of opposing each other, the
ground state of this 2D system is proposed to be an
Anderson-localized insulator despite the presence of strong electron
interaction.

\section{Experimental}
The experiments were performed on both single crystals and sintered
polycrystalline samples; the details of sample preparation were
described earlier in Ref.~\cite{fujimit}. All samples were single
phase as checked by x-ray diffraction, and the stoichiometries were
confirmed by energy-dispersive x-ray analysis in a scanning electron
microscope. Photoelectron spectroscopic experiments were carried out
in spectrometers that are equipped with Scienta analyzers, Gammadata
vacuum ultra-violet (VUV) lamps and monochromators to suppress the
satellite radiations. We have used the high-intensity He I VUV
radiation (h$\nu = 21.2$~eV) for these experiments. The total-system
energy resolutions were set at 8 meV full width at half maximum
(FWHM) for most of the measurements. The samples were cooled by
continuous-flow He cryostats and the temperature was controlled
within $\pm$ 0.5 K at any given temperature. The Fermi energy ($E_F$)
was determined at each temperature from spectra of polycrystalline Ag
in electrical contact with the sample. Single-crystalline samples
were primarily cleaved and used for angle-resolved band mapping at a
few selected temperatures, while scraped polycrystalline samples were
used to study in more detail the variation of the spectral weight at
$E_F$ as well as to check reproducibility in case of thermal cycling.

\section{Results and Discussion}
In order to address subtle temperature-dependent changes in the
electronic structure of any material with respect to the transport
properties, it is most relevant to monitor the spectral changes at
and near the Fermi energy, $E_F$. However, this requires a reliable
normalization procedure before different spectra can be compared. In
general, two distinct approaches have been adopted in the literature
for this purpose. In one approach, different spectra are scaled to
match at a given binding energy (BE); the specific BE (typically
0.5-0.6 eV) is normally chosen to be sufficiently removed from $E_F$,
such that subtle changes in the electronic structure of the system
that can be induced by temperature, are not expected to have any
effect in the vicinity of the chosen BE. In the other method, the
total integrated area under the spectrum over a certain energy window
is normalized, referring to the conservation of the number of
electrons in the system. In the present system, we find that both of
these approaches converge and lead to the same result, as illustrated
in the inset to Fig. 1a for the spectra of NiS$_2$ recorded at two
extreme temperatures of 20 K and 300 K. The spectra are normalized at
0.6-eV BE; interestingly, this normalization leads to the matching of
the two spectra down to 1.6 eV starting from 0.5-eV BE. This
extensive matching over the entire high-BE window ensures that the
two spectra shown in the inset of Fig. 1a have almost the same total
integrated areas. In fact, a normalization based on total integrated
areas leads to spectra essentially indistinguishable from those given
in the inset. It is also evident that there are no gross changes in
the spectra as a function of $T$~\cite{note}, which could influence
the normalization procedure.

In fact, temperature-dependent changes occur only close to $E_F$ over
a narrow energy range, as illustrated in the main panels of Figs. 1a
and 1b for single- and polycrystalline samples, respectively.
Interestingly, the data from single- and polycrystalline samples are
almost identical. In order to understand this observation, we have
carried out a detailed angle-resolved PE study of single-crystalline
samples (not shown here). While we found extensive dispersions of the
main intense valence band spectral region appearing at higher binding
energies, our angle-resolved measurements established a relative
insensitivity of the spectral features close to $E_F$ with respect to
the angle of detection. This is consistent with the striking
similarity of the spectral features from the single- and
polycrystalline samples.

In order to discuss the changes close to $E_F$ in more detail, we
show in Fig. 2 (main panel) a set of representative spectra covering
a narrow energy scale at various temperatures. Far below $E_F$ (BE$>$
0.5-eV), the spectra are essentially identical, as already
illustrated in Fig. 1 for the two extreme temperatures. In the BE
region between 100 and 500 meV, however, there is a systematic,
though small, depletion of spectral weight with increasing
temperature that compensates for the increase in spectral weight
above $E_F$ with $T$. Here, we focus on the electronic states close
to $E_F$, responsible for the transport properties, which exhibit
remarkable changes with temperature. It is well known that in the
case of a metal the spectral weight at and near $E_F$ changes
significantly with temperature, which is easily understood in terms
of the Fermi-Dirac statistics. The changes observed here, however,
are distinctly non-Fermi-Dirac type, as we shall show below by a more
detailed analysis.

Already without recourse to a detailed analysis, several important
features can be recognized, which are particularly interesting in
connection with the transport data reported for NiS$_2$. On the basis
of the shown PE spectra, there is compelling evidence for the surface
of NiS$_2$ to be metallic at room temperature. These evidences are:
(i) There is a large spectral weight at $E_F$ in the room-temperature
spectrum. (ii) There is a continuous and substantial spectral weight
up to about 100 meV above $E_F$ in the same spectrum. (iii) The
spectral weight {\em above} $E_F$ shows a characteristic Fermi-Dirac
type depletion of spectral intensity with decreasing temperature,
spread over the expected energy scale related to the thermal energy.
In view of the total experimental energy resolution of the present PE
experiments of 8 meV (FWHM), none of these observations can be
explained by broadening due to finite resolution. The observations
are clearly incompatible with the optical data that show NiS$_2$ to
be a wide-gap insulator. On the other hand, the results are
consistent with the Hall-effect data~\cite{thio1} and suggest that
the surface of NiS$_2$ is metallic, particularly when considering the
high surface sensitivity of PE with a mean sampling depth of
$\approx$ 7~{\AA}. Thus, the surface layer of NiS$_2$, only a few
unit cells thick, has metallic character near room temperature, while
the bulk of the sample, inaccessible to PE, remains an insulator.

It is interesting to note that the variation of the spectral function
with temperature, shown on the main panel of Fig.~2, is very
different from that of a normal metal. This can be inferred from the
inset that displays corresponding spectra of polycrystalline Ag metal
taken with the same setup. Here, the observed $T$-changes in the
spectral function of Ag are readily understandable in terms of
Fermi-Dirac (FD) statistics alone. With an essentially constant
density of states
 in the vicinity of $E_F$ (${\cal D}(E_F)$), this causes the spectrum to
recover the weight lost {\em above} $E_F$ with decreasing temperature
almost in a symmetrical manner and immediately {\em below} $E_F$,
thus preserving the total spectral weight. Another consequence of
$\cal D$ being unchanged with $T$ is that all spectra of Ag go
through a common point at $E_F$ in spite of changing $T$, since the
Fermi-Dirac statistics does not affect the spectral weight {\em at}
the Fermi energy. An almost identical behavior of the spectral
function is also observed for NiS (see second inset in Fig. 2). In
contrast, the behavior of the PE spectra of NiS$_2$ is qualitatively
different. In particular, the redistribution of the spectral function
{\em below} $E_F$ takes place over a much wider energy range than one
would expect explicitly on a thermal scale, while {\em above} $E_F$
it appears to be controlled by $T$. Hence, the changes in the
spectral function of NiS$_2$ cannot be understood in terms of a {\it
fixed} $\cal D$ around $E_F$ and  FD statistics, in contrast to
normal metals. Instead, we are forced to conclude from the raw data
that the single-particle excitation spectrum of the surface of
NiS$_2$ is characterized by a temperature dependence of ${\cal D}$
itself.

Due to the presence of thermal broadening and resolution broadening,
a quantitative estimate of ${\cal D}(E_F)$ at various temperatures
cannot be obtained without recourse to simulations of the spectra in
terms of a model ${\cal D} (E)$. With the high energy resolution
achieved in the present PE experiments, thermal broadening is by far
the dominant effect. This contribution, however, can be readily
removed from the spectra by assuming a symmetric ${\cal D} (E)
$~\cite{mohit}, or by dividing the raw spectra by the FD distribution
function
\cite{susaki}. We illustrate the results of the latter analysis by
the inset in Fig. 3, demonstrating a remarkable depletion of spectral
weight with decreasing temperature over a wide energy range, with the
strongest effects at $E_F$. As pointed out further above, the
corresponding amount of spectral weight is recovered almost uniformly
distributed over the BE range from 0.15~eV to 0.45~eV, with the
effects of FD distribution being virtually absent at such high BE.

Besides these two methods, the $T$ dependence of ${\cal D} (E_F)$ was
also obtained by direct fitting of the spectra of Fig.~2 with a
${\cal D}(E)$ described by a polynomial function, multiplied by the
FD distribution at a given $T$ and convoluted by a Gaussian for the
known resolution. All these different analyses result in similar
values of ${\cal D} (E_F)$, demonstrating an insensitivity on details
of the model. The resulting ${\cal D} (E_F)$ are plotted on the main
panel of Fig.~3 as a function of $T$, with error bars that contain
the variations of ${\cal D} (E_F)$ from the different analyses.
${\cal D} (E_F)$ increases slightly with decreasing temperature from
297~K to 260~K, and then decreases progressively down to 115~K.
Between 115~K and 75~K, there is a more pronounced decrease in ${\cal
D} (E_F)$, suggesting the opening of a gap-like structure in the
single-particle excitation spectrum; below 75~K, the gap-like
structure is fully developed and  ${\cal D} (E_F)$ remains
essentially unchanged. The formation of an energy gap in the
electronic structure of NiS$_2$ thus occurs in the very temperature
range, where the resistivity is known to increase~\cite{honigres}.

This behavior is not specific to NiS$_2$ alone. As mentioned before,
also the related compounds NiS$_{2-x}$Se$_x$ are bulk insulators,
with the resistivity data suggesting a 2D metallic overlayer for $x
\le$~0.4~\cite{honigres}. PE spectra of NiS$_{2-x}$Se$_x$, with $x
= 0.3$ and $x = 0.4$, were recorded at various temperatures and
reveal a behavior similar to that of NiS$_2$. Data for
NiS$_{1.6}$Se$_{0.4}$ are included in Fig.~3, and they clearly
resemble the behavior of NiS$_2$, with a tiny initial increase in
${\cal D} (E_F)$ between 297 and 260~K, followed by a moderate
decrease down to 75~K, and then the signature of a gap at $\approx$
35~K. The PE results for NiS$_{1.7}$Se$_{0.3}$ (not shown here) are
again similar, with the formation of a gap-like structure at
$\approx$~55~K. Interestingly, the gap-like structures in the
single-particle excitation spectrum form at progressively lower
temperatures with increasing $x$, i.e., at about 75~K, 55~K, and 35~K
for $x=0$, 0.3, and 0.4, respectively. This follows the trend of
decreasing $T$, at which the upturn in resistivity has been observed
for NiS$_{2-x}$Se$_x$
\cite{honigres}, and establishes a close relationship between the
transport properties and the observed decrease of ${\cal D}(E_F)$.

The details of the temperature-dependent surface electronic
structure, as derived from the PE spectra of NiS$_2$ and related
compounds, further suggest several important implications. We first
note that ${\cal D}(E_F)$ for any of the three compounds studied, in
spite of the pronounced gap-like structure, does not vanish
completely, even at the lowest $T$. While the gap-like structure in
${\cal D}(E)$ at the low-$T$ limit is presumably driven by strong
correlation effects, the persistence of a finite ${\cal D}(E_F)$
suggests that the ground state is not a Mott insulator with a fully
developed energy gap; it rather indicates that Anderson localization
driven by disorder is the origin of the insulating behavior.

The overall dependence of ${\cal D} $ on $T$ displayed in Fig.~3
shows that the gap-like structure disappears over a relatively narrow
temperature range of $\approx 40$~K rather than by a gradual filling
of the gap over a larger $T$ interval. This cannot be explained by
thermal excitations of charge carriers alone implying that the
underlying electronic structure itself changes rapidly, accompanying
the metal-insulator crossover over a narrow $T$ range. This behavior
might be related to a temperature-dependent screening of correlation
effects by itinerant electrons. At high temperatures the screening is
highly effective, leading to a less correlated state and consequently
to the disappearance of the gap-like structure in the single-particle
excitation spectra. At lower temperatures, however, Anderson
localization of the electrons leads to less effective screening, with
a gap forming in the excitation spectrum. While this may possibly be
a continuous change-over with temperature, as suggested by the slow
change of ${\cal D}(E_F)$ at higher $T$, it is most remarkable in a
narrow temperature interval, where a change in the transport
properties was observed. If indeed such a screening mechanism is
relevant, it has to have a non-trivial dependence on $T$ beyond the
thermal excitation of mobile charge carriers.

At this point, it is tempting to rationalize the presence of
metal-like surface layers in these systems. One obvious possibility
is a deviation of the surface stoichiometry from the bulk, leading to
a doping of the surface layer with charge carriers. However, the
large ${\cal D}(E_F)$ observed near room temperature would correspond
to a very high doping level and hence to a significant deviation from
stoichiometry. This is incompatible with the intensity ratios of
core-level PE spectra from Ni, S, and Se, which were found to follow
the nominal bulk compositions in all samples. Moreover, if the doped
charge carriers arising from non-stoichiometry were indeed mobile,
without altering the underlying electronic structure, there would be
no reason for the charge carriers to reside only in the surface
layer.

A possible explanation is obtained from the unusual properties of
NiS$_2$ compared to CoS$_2$ and CuS$_2$. While NiS$_2$ is a bulk
insulator, the latter two are bulk metals~\cite{wil}. The exceptional
behavior of NiS$_2$ shows up also in the lattice parameters being
{\em larger} than expected from their systematic variation in the
pyrite family, $M$S$_2$, with $M$ = Mn-Zn~\cite{wil}. The unusually
expanded lattice of NiS$_2$ is expected to lead to a reduction of the
bandwidth, driving bulk NiS$_2$ insulating, in contrast to the
metallic ground states of CoS$_2$ and CuS$_2$. This point of view is
supported by the fact that bulk NiS$_2$ turns metallic at a pressure
of 46 kbar~\cite{mori}. Given this relation between the lattice
parameters and the metallic character of bulk NiS$_2$, a reduction of
the lattice parameters at the surface could lead to an altered
surface electronic structure of NiS$_2$ with the tendency towards a
more metallic character. And in fact, preliminary {\it ab initio}
full-potential calculations of the surface electronic structure of
NiS$_2$~\cite{unpub} indicate a reduction of the lattice parameters
and consequently an increase in $d$ bandwidth at the surface for
several crystallographic surfaces.

To summarize, we propose the following scenario: Lattice relaxation
near the surface region of NiS$_2$ and related compounds tends to
drive the surface metallic, which, however, is inhibited by the 2D
nature of the surface layer; instead Anderson localization leads to
an insulating ground state of the surface layer. At higher
temperatures, crossover towards a metallic behavior occurs, as
evidenced by the transport properties. This crossover is
characterized by the disappearance of the gap-like feature in the
electronic structure around $E_F$ and by a rapid increase in ${\cal
D}(E_F)$, driven by a decrease of the effective Coulomb interaction
strength as well as of the disorder potential due to screening by a
larger number of thermally excited electrons. However, the narrow
temperature range, in which this strong change in the electronic
structure occurs, is remarkable and requires further theoretical
explanation.

The authors thank S. Das Sarma, H.R. Krishnamurthy, and P. Mahadevan
for fruitful discussions. The work was supported by the Deutsche
Forschungsgemeinschaft, Sfb-290, TPA 06, and the Deutsches Zentrum
f{\"u}r Luft- und Raumfahrt e.V., Project INI-012-99, as well as by
the Department of Science and Technology and Board of Research in
Nuclear Sciences, Government of India. D.D.S. thanks the Freie
Universit{\"a}t Berlin and the University of Tokyo for hospitality.
C.M. acknowledges financial support by the Alexander von Humboldt
Foundation.


\section*{Figure Captions}

\noindent
Fig.~1. PE spectra of NiS$_2$ from (a) single-crystalline and (b)
polycrystalline samples at the given temperature. The inset
illustrates the normalization procedure over a larger range of
binding energies.
\\

\noindent
Fig.~2. PE spectra of polycrystalline NiS$_2$, recorded at various
temperatures. Insets: corresponding spectra for polycrystalline Ag
and polycrystalline NiS. \\

\noindent
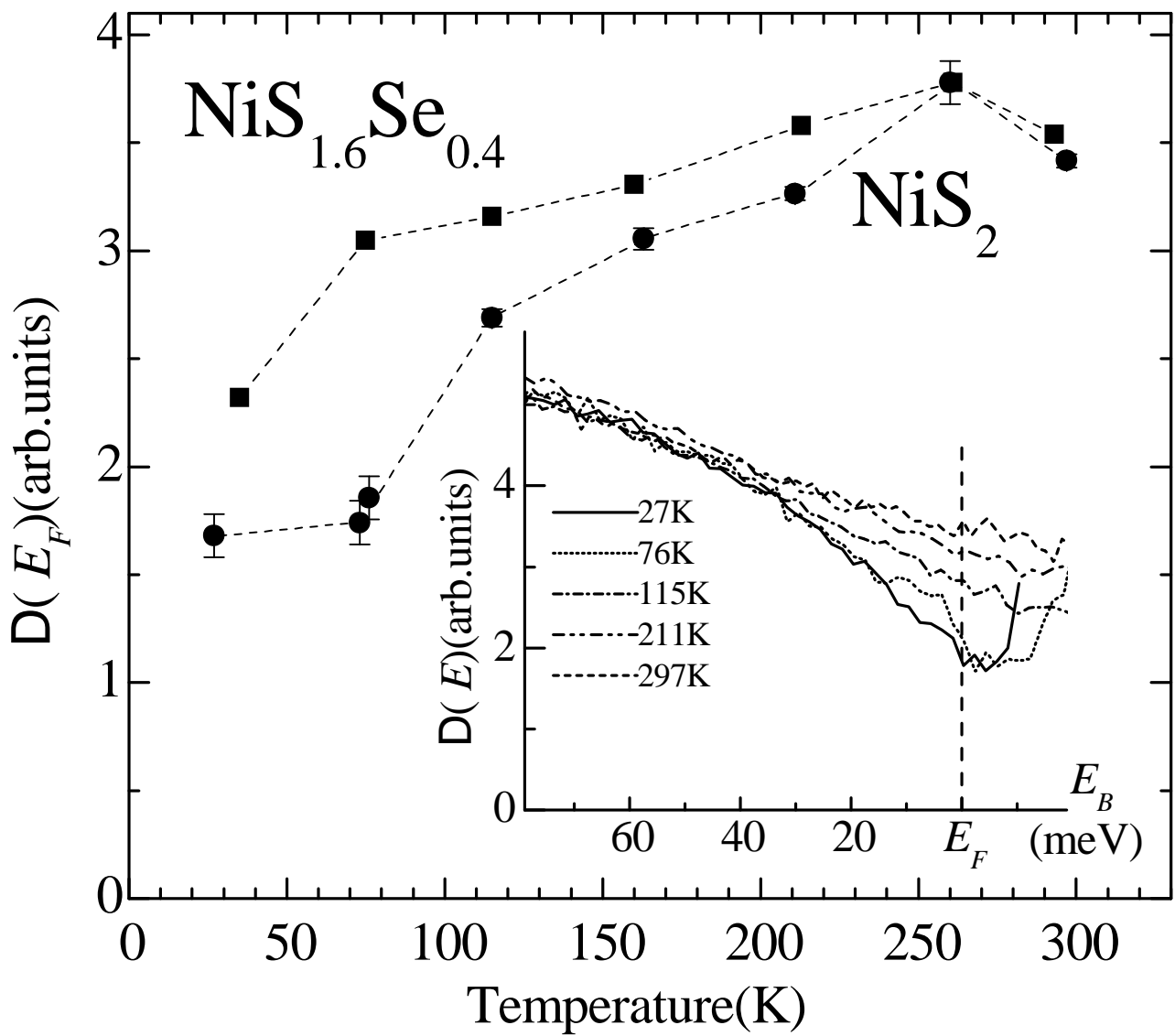
Fig.~3. $T$ dependence of ${\cal D} (E_F)$ of NiS$_2$, extracted from
the spectra of Fig.~2; data for NiS$_{1.6}$Se$_{0.4}$ are included.
Inset: ${\cal D} (E)$ at various temperatures, obtained as described
in the text.


\begin{thebibliography}{widest-label}
\bibitem{note} One change that could have been expected is a reduction
in the phonon broadening of the spectral features. This contribution,
however, is obviously so small as compared to the overall width of
the observed feature, that it is not noticeable at all.
\end{thebibliography}
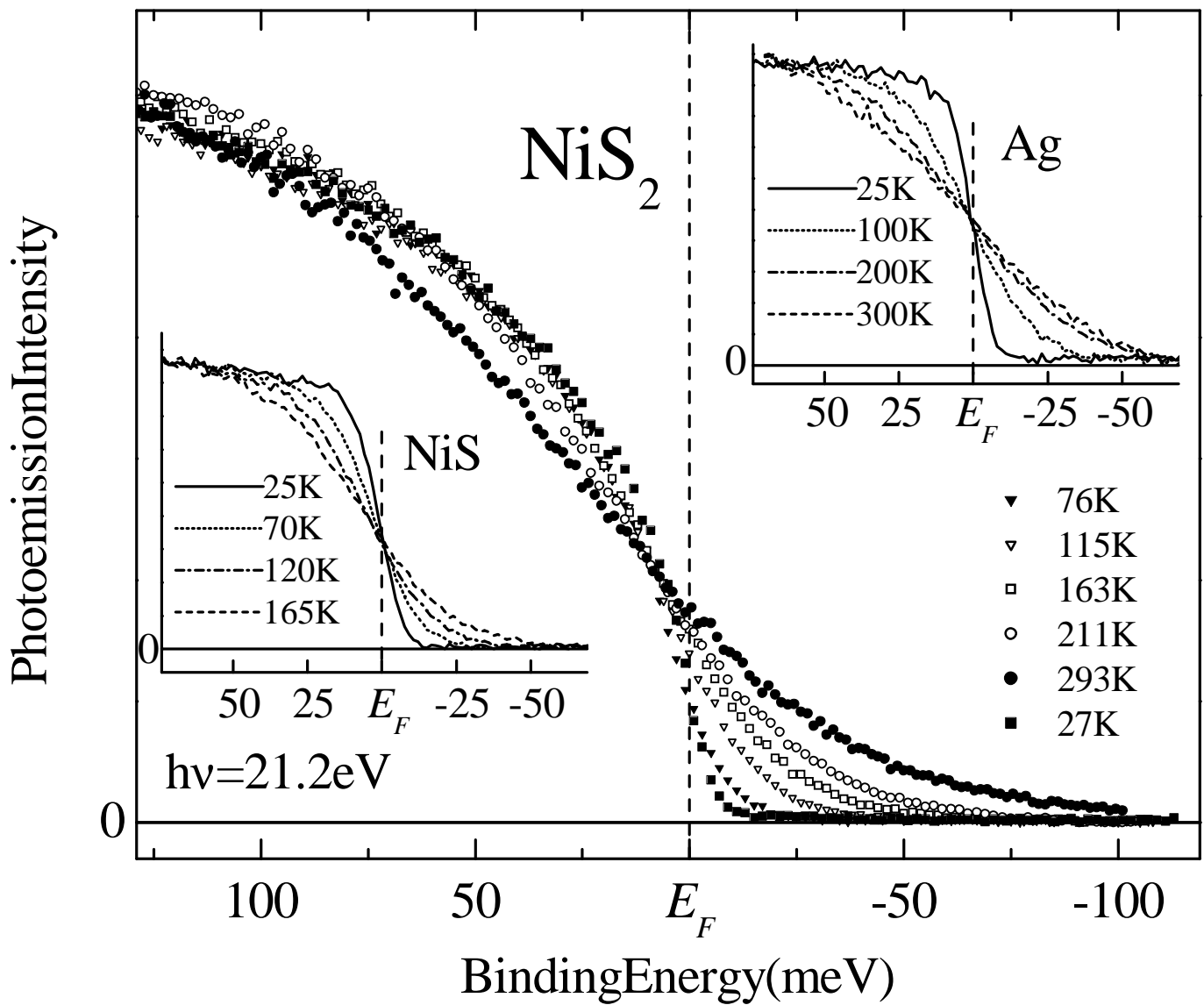
\end{document}